\documentclass[preprint,showpacs,showkeys,preprintnumbers,amsmath,amssymb,pre]{revtex4-1}
\usepackage[final]{graphicx} 
\usepackage{dcolumn}
\usepackage{bm}
\usepackage{array}

\newcommand\Rey{\mbox{\textit{Re}}}  
\newcommand\etal{\mbox{\textit{et al.}}}

\graphicspath{{./fig/}}

\begin{document}

\title{Stability and angular-momentum transport of fluid flows between
  corotating cylinders}

\author{M. Avila}

\affiliation{Max Planck Institute for Dynamics and Self-Organization
  (MPIDS), 37077 G\"ottingen, Germany\\ Institute of Fluid Mechanics,
  Friedrich-Alexander-Universit\"at Erlangen-N\"urnberg,
  Cauerstra{\ss}e 4, 91058 Erlangen, Germany}

\date{\today}

\begin{abstract}
  Turbulent transport of angular momentum is a necessary process to
  explain accretion in astrophysical disks. Although the hydrodynamic
  stability of disk-like flows has been tested in experiments, results
  are contradictory and suggest either laminar or turbulent
  flow. Direct numerical simulations reported here show that currently
  investigated laboratory flows are hydrodynamically unstable and
  become turbulent at low Reynolds numbers. The underlying
  instabilities stem from the axial boundary conditions, affect the
  flow globally and enhance angular momentum transport.
\end{abstract}


\maketitle 

Accretion in astrophysical disks requires the flow of mass towards a
central gravitating body. The ensuing loss of momentum must be
balanced by outward angular momentum transfer among gas particles
\cite{pringle1981}. If the motion of orbiting gas was laminar
molecular transport would be orders of magnitude too slow for
accretion to take place, and so considering a turbulent viscosity
becomes necessary \cite{shakura1973}. However, in Keplerian disks the
gas rotates as $\Omega \propto r^{-3/2}$ and laminar motion is
linearly stable according to the Rayleigh criterion. Although axial
magnetic fields can drive turbulence via the magnetorotational
instability \cite{balbus1998,*balbus2003}, it is not clear whether
this operates in weakly ionized disks. On the other hand, it is well
known that linearly stable shear flows (such as pipe flow) can become
turbulent due to finite amplitude disturbances. Whether Keplerian
flows are susceptible to such a transition scenario or remain stable
despite the large Reynolds numbers (\Rey), is a topic of great
interest and the source of much controversy
\cite{balbus2011}.

The stability of disk-like flows is typically probed in laboratory
experiments of fluid between two concentric and independently rotating
cylinders, Taylor--Couette flow (TCF). In the infinite-cylinder
idealization, the Navier--Stokes equations admit a pure rotary
solution
\begin{equation}\label{eq:couette}
  \Omega(r) = \dfrac{\Omega_2r_2^2-\Omega_1r_1^2}{r_2^2-r_1^2} +
  \dfrac{(\Omega_1-\Omega_2)(r_1r_2)^2}{r_2^2-r_1^2}\,\dfrac{1}{r^2},
\end{equation}
commonly referred to as Couette flow. Here $r_1$ and $r_2$ are the
radii of the inner and outer cylinders and $\Omega_1$ and $\Omega_2$
their angular velocities. When $(r_1/r_2)^2<\Omega_2/\Omega_1<1$ the
angular velocity decreases outward but the angular momentum increases
(quasi-Keplerian flows). Accretion disks are stratified in the axial
direction and are thus best modelled considering an unbounded domain,
thereby avoiding artificial boundary conditions
\cite{barranco2005}. Most experiments and simulations focus, however,
on the physics of the disk's midplane and neglect
stratification. Under this assumption simulations typically employ
axially periodic boundary conditions, whereas in experiments cylinders
have a finite-length $h$. Hence the degree to which \eqref{eq:couette}
may be approximated is compromised by the axial boundary conditions
and length-to-gap aspect-ratio $\Gamma=h/(r_2-r_1)$. In particular,
solid axial boundaries result in a basic state with nonzero radial and
axial velocity components (Ekman flow). Hence, producing Couette-like
profiles in experiments poses an extraordinary challenge, which may be
addressed by considering very tall cylinders or splitting the endwalls
in several rings that rotate at independent angular speeds
\cite{hollerbach2004}. The latter strategy has been implemented in the
Princeton Taylor--Couette experiment \cite{schartman2009}, which has a
short aspect-ratio $\Gamma=2.104$ but whose endwalls are split into
two independently rotating rings. From simultaneous Laser Doppler
Velocity measurements of azimuthal and radial velocity components, Ji
\etal\ \cite{ji2006} suggest that quasi-Keplerian flows at Reynolds
numbers up to millions are essentially laminar. From this, they
conclude that purely hydrodynamic mechanisms cannot transport angular
momentum at the rates required for accretion to occur in disks.

This conclusion has been recently challenged in a new experimental
study by Paoletti and Lathrop \cite{paoletti2011}, who report from
direct torque measurements at the inner cylinder that Keplerian flows
at $\Rey\gtrsim10^6$ are fully turbulent. When extrapolated to
astrophysical disks, their results indicate that transport occurs at
accretion relevant rates, in agreement with previous estimations
\cite{richard1999}. Despite having tall cylinders $\Gamma=11.47$, the
Maryland experiment has solid endwalls that are attached to the outer
cylinder and hence cannot be rotated independently. Although these
boundary conditions are known to generate vigorous Ekman vortices and
greatly increase the exerted torque, their contribution is discarded
by dividing the inner cylinder into three sections and measuring
torque only in the central one. Despite efforts in the Princeton and
Maryland experimental setups to mitigate endwall effects it is,
however, unclear whether their results can be used to infer the
stability of flows in astrophysical disks \cite{balbus2011}.  In this
\emph{Letter} it is shown that current laboratory experiments of
quasi-Keplerian flows become turbulent already at $\Rey=O(10^3)$ due
to hydrodynamic instabilities stemming from the axial boundary
conditions. Moreover, it is found that turbulence fills the entire
flow domain and as a result the momentum transfer is globally
enhanced.

Here direct numerical simulations of flows with the precise geometry
and boundary conditions of the Princeton and Maryland experiments were
performed. The Navier--Stokes equations for an incompressible
Newtonian fluid of velocity $\mathbf{v}$
\begin{equation}\label{eq:NS}
  \partial_{t}\mathbf{v} + (\mathbf{v}\cdot\nabla)\mathbf{v} = -\nabla
  p + \Delta\mathbf{v},\quad \nabla\cdot\mathbf{v} = 0 \;,
\end{equation}
were rendered dimensionless by scaling lengths and time with the
gap-width $d=r_2-r_1$ and viscous time $d^2/\nu$, where $\nu$ is the
kinematic viscosity of the fluid. The solution of \eqref{eq:NS} was
formulated in primitive variables in cylindrical coordinates
$(r,\theta,z)$ and a second-order time-splitting method with
consistent boundary conditions for the pressure was used
\cite{hughes1998,*mercader2010}. The spatial discretization consists
of Chebyshev collocation in $(r,z)$ and a Galerkin-Fourier expansion
in $\theta$.  The code converges spectrally in the three directions
\cite{avila2008} and was validated against a Legendre-Fourier-Galerkin
code \cite{marques2006}. Here the resolution was chosen to ensure that
computed torque values were accurate to at least 1\%.

\begin{figure}
  \centering
  \includegraphics[width=0.6\linewidth]{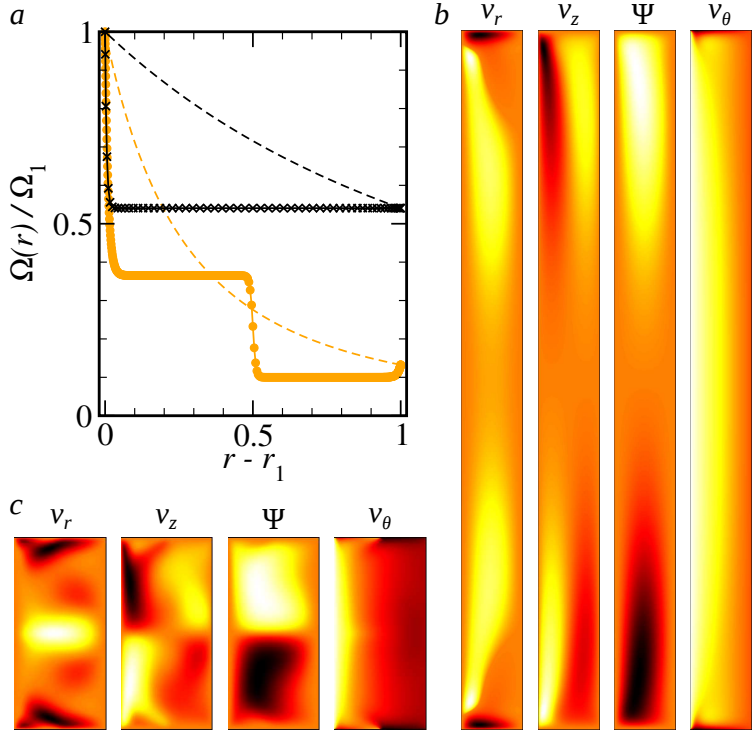}
  \caption{(color online) ($a$) Angular velocity at the endwalls for
    the Princeton (circles) and Maryland (crosses) experiments, and
    Couette flow (dashed lines). ($b$--$c$) Steady basic states at
    $\Rey=320$ (Maryland) and $\Rey=772$ (Princeton). White (black)
    corresponds to maximum (minimum) velocity and the radial direction
    is horizontal with left (right) corresponding to inner (outer)
    cylinder.}
  \label{fig:bflow}
\end{figure}

The geometry of the Taylor--Couette system is specified by the
radii-ratio $\eta=r_1/r_2$ and the length-to-gap aspect-ratio
$\Gamma$. The dimensionless boundary conditions at the cylinders read
$(v_r,v_\theta,v_z)[r_{1,2},\theta,z]=(0,Re_{1,2},0),$
where $Re_1=d r_1\Omega_1/\nu$ ($Re_2=d r_2\Omega_2/\nu$) is the inner
(outer) cylinder Reynolds number.  Because of differential rotation
the angular velocity changes abruptly at adjacent rotating
boundaries. In the Princeton experiment, the endwall is split at
mid-radius $r_m=(r_1+r_2)/2$ into two independently rotating
rings. Hence, there are four independent angular speeds; $\Omega_1$
and $\Omega_2$, for inner and outer cylinder, and $\Omega_3$ and
$\Omega_4$, for inner and outer rings. To preserve spectral
convergence discontinuities in angular velocity were regularized,
yielding the following boundary condition at the endwalls $z=\pm
\Gamma/2$
\begin{align*}
  \Omega(r) = & (\Omega_1-\Omega_3)\exp[-(r-r_1)/\epsilon]+\\ &
  (\Omega_2-\Omega_4)\exp[-(r_2-r)/\epsilon] +\\ &
  \dfrac{\Omega_3+\Omega_4}{2} +
  \dfrac{\Omega_4-\Omega_3}{2}\tanh[(r-r_m)/\epsilon],
\end{align*}
with $\epsilon\in[5\times10^{-3},10^{-2}]$ (see
Ref.~\onlinecite{lopez1998}). The boundary condition modeling the
Princeton experiment \cite{ji2006} is shown as circles in
Fig.~\ref{fig:bflow}$a$. Due to the sharp gradient
$\partial\Omega/\partial r|_{r_m}$ and the clustering of Chebyshev
points close to the boundaries, a large number of radial points
($n_r=351$) was required to accurately simulate the split endwall. In
the axial and azimuthal directions up to $n_z=281$ Chebyshev points
and $n_\theta=256$ Fourier modes were used. The Maryland experiment
has a single solid ring attached to the outer cylinder
($\Omega_4=\Omega_3=\Omega_2$) and there is only a strong gradient at
$r_1$ (see crosses in Fig.~\ref{fig:bflow}$a$). Here up to $n_z=601$,
$n_\theta=384$ and $n_r=61$ were used.

To put TCF in the wider context of rotating shear flows it is useful
to define a shear Reynolds number $\Rey= 2/(1+\eta)|Re_2\eta -Re_1|$
and a rotation number $R_\Omega= (1-\eta)(Re_1+Re_2)/(Re_2\eta
-Re_1)$, which measure the ratio of shear to viscous forces and the
ratio of mean rotation to shear \cite{dubrulle2005},
respectively. Here the sign of $R_\Omega$ distinguishes between
cyclonic ($R_\Omega>0$) and anticyclonic flows ($R_\Omega<0$), with
$-2<R_{\Omega}<-1$ corresponding to quasi-Keplerian rotation. In the
experiments of Ji \etal\ \cite{ji2006} $R_\Omega=-1.038$ and
$\eta=0.3478$, and the same values were used here, whereas Lathrop and
Paoletti \cite{paoletti2011} have systematically studied both cyclonic
and anticyclonic regimes at $\eta=0.7245$. Here, $\eta=0.7245$ and the
value $R_\Omega=-1.047$ was chosen (corresponding to their Rossby
number $Ro=Re_1/(\eta Re_2)-1=0.85$).

The endwall influence in the Maryland experiment is illustrated in
Fig.~\ref{fig:bflow}$b$, showing the velocity field of the steady
basic state at $\Rey=320$. At the endwalls there is a strong negative
radial velocity inflow, which generates axial velocities pointing
towards mid-height along the inner cylinder and result in an axially
dependent azimuthal velocity. Figure~\ref{fig:bflow}$c$ shows the
basic state of the Princeton experiment at $\Rey=772$. Because of the
small aspect-ratio the meridional circulation generates a strong
radial outward flow at mid-height that increases the outward transport
of azimuthal velocity. Were the angular speeds of the endwall rings
selected according to the ideal Couette profile
\cite{hollerbach2004,kageyama2004} instead of the values used in
experiments \cite{ji2006} and reproduced here, profiles with weaker
meridional circulation and hence closer to Couette flow could be
obtained. This approach was used in previous numerical simulations of
the Princeton configuration \cite{obabko2008}.

\begin{figure}
  \centering
  \includegraphics[width=0.6\linewidth,clip=]{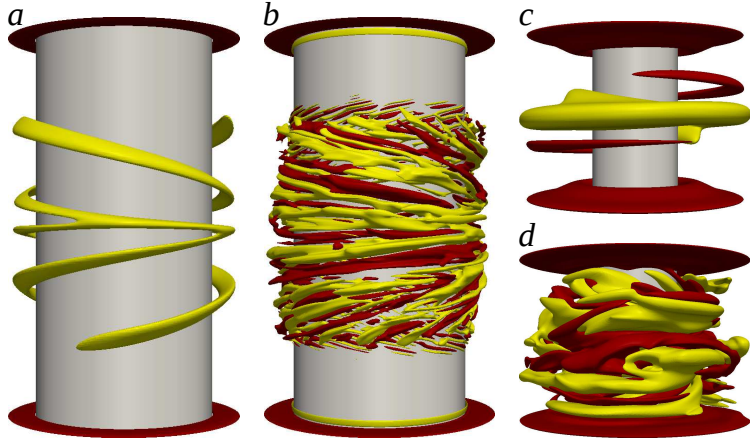}
  \caption{(color online) Three-dimensional view of isosurfaces of
    negative (red) and positive (yellow) radial velocity. ($a$) $m=2$
    rotating wave at $\Rey=1332$ and ($b$) turbulent flow at
    $\Rey=5328$ from simulations of the Maryland experiment. ($c$)
    Modulated rotating wave with $m=1$ and $m=2$ at $\Rey=1545$ and
    ($d$) turbulent flow at $\Rey=6437$ from simulations of the
    Princeton experiment.}
  \label{fig:inst}
\end{figure}

The visualizations of Fig.~\ref{fig:bflow}$b$--$c$ hint at the
difficulty of realizing quasi-Keplerian profiles in a laboratory
experiment even at very low Reynolds numbers. The endwall boundary
conditions change the velocity field not just locally but globally
across the domain. In fact, at slightly higher Reynolds numbers the
flow becomes three-dimensional and time-dependent via supercritical
Hopf bifurcations. In the Maryland configuration instability occurs
first at $\Rey_c=352$ to prograde rotating waves with azimuthal
wavenumber $m=5$ and localized at the endwalls. Beyond $\Rey_c$
multiplicity of states, with different symmetries and wavenumber
$m\in[2,5]$, was found, whereas for $\Rey \gtrsim 1330$ only a global
$m=2$ mode remained stable and was obtained regardless of initial
conditions (see Fig.~\ref{fig:inst}$a$). Further increasing the
Reynolds number led to modulated waves and a quick transition to
temporal chaos at about $\Rey\simeq 1600$. Subsequently, spatial
periodicity was lost and the spectra broadens as the flow became
gradually turbulent; Fig.~\ref{fig:inst}$b$ shows a flow snapshot at
$\Rey=5328$. This transition picture is also representative of the
simulations of the Princeton experiment. Here the basic steady state
becomes unstable at $\Rey=1448$ almost simultaneously to $m=1$ and
$m=2$ rotating waves, which were found to coexist in space and time
(see Fig.~\ref{fig:inst}$c$). By no means is this situation generic:
changing the relative rotation of the cylinders one of $m=1,2$ was
found to bifurcate first. Further increasing the Reynolds number led
to very complex and strongly three-dimensional flow as shown in
Fig.~\ref{fig:inst}$d$.

\begin{figure}
  \centering
  \includegraphics[width=0.6\linewidth,clip=]{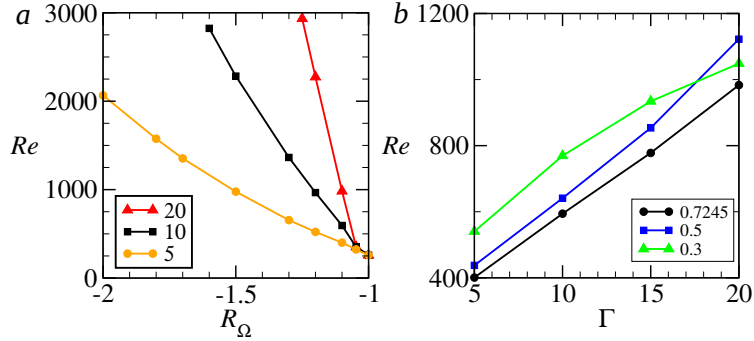}
  \caption{(color online) Stability curves of quasi-Keplerian
    Taylor--Couette flows with endwalls attached to the outer
    cylinder: ($a$) $\eta=0.7245$ and $\Gamma$ as in the legend, ($b$)
    $R_\Omega=-1.1$ and $\eta$ as in the legend.}
  \label{fig:stab}
\end{figure}

The stability of quasi-Keplerian TCF with endwalls attached to the
outer cylinder and $\eta=0.7245$ is shown in
Fig.~\ref{fig:stab}$a$. The minimum critical Reynolds number is
attained at the Rayleigh line ($R_\Omega=-1$) and increases as
differential rotation decreases, but with $Re_c<10^4$ across the whole
quasi-Keplerian regime. Close to the Rayleigh line modes localized to
the endwalls bifurcate first, whereas for $R_\Omega\lesssim -1.1$
global instability modes (as in Fig.~\ref{fig:inst}$a$) dominate. The
former are similar to those observed experimentally in
Ref.~\onlinecite{andereck1986} and the latter are similar to those
reported by Avila \etal\ \cite{avila2008}, who studied global boundary
layer effects on flows between exactly co-rotating cylinders and
stationary endwalls. Figure~\ref{fig:stab}$b$ further shows that
endwall instabilities depend weakly on geometry and hence generically
govern the dynamics of quasi-Keplerian TCF. It is worth noting that
endwall instabilities persist beyond the Rayleigh line and coexist
with Taylor vortices close to the onset of centrifugal instability.

\begin{figure}
  \centering
  \includegraphics[width=0.6\linewidth,clip=]{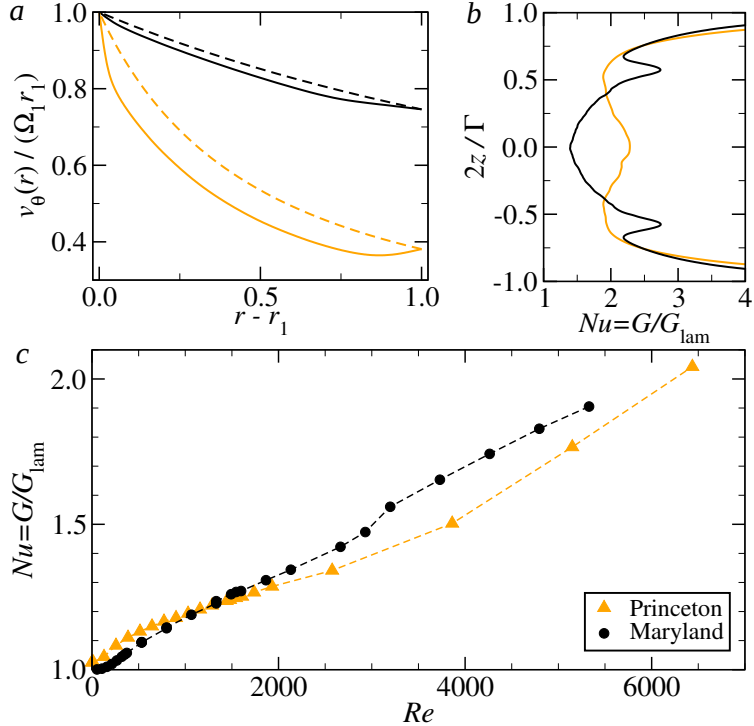}
  \caption{(color online) ($a$) Average azimuthal velocity at
    mid-height for simulations of the Maryland ($\Rey=5328$, black
    solid line) and Princeton ($\Rey=6437$, orange solid line)
    experiments. The dashed lines are laminar Couette flow. ($b$)
    Torque $Nu=G/G_{\text{lam}}$ along the inner cylinder, solid lines
    as in ($a$). ($c$) Reynolds number dependence of the torque on the
    central section of the inner cylinder ($0.6>|2z/\Gamma|$ and
    $0.4>|2z/\Gamma|$ for simulations of the Maryland and Princeton
    experiments).}
  \label{fig:torque}
\end{figure}

The onset of hydrodynamic instability and transition to turbulence are
expected to radically change the radial transport of azimuthal
momentum. The solid lines in Fig.~\ref{fig:torque}$a$ show normalized
average azimuthal velocity profiles $\langle
v_\theta\rangle_{\theta,t}/(r_1\Omega_1)$ at mid-height for
simulations of the Maryland experiment at $\Rey=5328$ (black solid
line) and Princeton experiment at $\Rey=6437$ (gray solid line, orange
online). At the inner cylinder the profiles are steeper than laminar
Couette flow (dashed lines), implying larger torques on the cylinder
surface.  It is worth noting that in TCF between infinite cylinders
the transverse current of azimuthal motion $ J^\Omega =
r^3\big[\langle v_r\Omega\rangle_{\theta,z,t} -
  \nu\partial_r\langle\Omega\rangle_{\theta,z,t}\big] $ is a conserved
quantity \cite{eckhardt2007}, and as a consequence the dimensionless
torque $G=\nu^{-2}J^\Omega$ is the same at inner and outer cylinder.
This does not hold, however, for flows confined by no-slip axial
boundaries. Torque profiles along the inner cylinder, normalized with
the laminar Couette torque, are shown in
Figure~\ref{fig:torque}$b$. Because of the sharp change in $\Omega$
occurring across a small gap between inner cylinder and endwalls (see
Fig.~\ref{fig:bflow}$a$), the torque required to rotate the inner
cylinder faster than the endwall is very large. This increase in local
torque as the endwalls $2z/\Gamma=\pm1$ are approached can be seen in
Figure~\ref{fig:torque}$b$. Although the direct contribution of the
endwall is largely avoided by the measurement technique in the
experiments, the torque in the central section remains well above
laminar because of turbulent fluctuations.

The simulations of the Maryland experiment show a clear change in the
torque behavior at about $\Rey=3000$ (see black curve in
Fig.~\ref{fig:torque}$c$). This is related to the appearance of the
two torque peaks at $2z/\Gamma\sim\pm0.5$ in Fig.~\ref{fig:torque}$b$
and is caused by the onset of small-scale vortices which have opposite
spiral orientation to the structure of the primary rotating wave (see
Fig.\ref{fig:inst}$b$). In both experiments the torque has already
doubled the laminar value at $\Rey\sim6000$ due to the endwall driven
instabilities.

In spite of the disparity in Reynolds numbers, it is tempting to
compare these numerical results to experimental observations. Van Gils
\etal\ \cite{vangils2011} have measured torque at $\Rey>10^5$ and have
observed an effective universal scaling law
$Nu=G/G_{\text{lam}}=a(R_\Omega)\Rey^{0.76}$ that holds throughout the
linearly unstable regime of TCF. Surprisingly, Paoletti and Lathrop
\cite{paoletti2011} (see also Ref.~\onlinecite{paoletti2012}) have
demonstrated that this law applies to linearly stable cyclonic and
anti-cyclonic Rayleigh-stable regimes as well, that is including
quasi-Keplerian rotation. It is then natural to ask how this universal
behavior connects to the complex flows uncovered in this work. It is
speculated here that a transition between endwall-driven turbulence to
the universal $Nu \propto \Rey^{0.76}$ scaling reported in experiments
may take place at intermediate \Rey. Interestingly, in the case of
stationary outer cylinder a cross-over marking the transition from
centrifugally to shear-driven turbulence at $\Rey\simeq13000$ was
reported \cite{lathrop1992,*lewis1999}. If an analogous cross-over was
found in quasi-Keplerian flows and shown to be independent of
aspect-ratio and endwall boundary condition, a strong case for the
existence of hydrodynamic turbulence in astrophysical disks would be
made. In fact, ingredients of shear-driven turbulence such as
transient growth of disturbances are found also in quasi-Keplerian
flows, although significantly only at $\Rey=O(10^6)$ \cite{yecko2004}.
On the other hand, it would be interesting to investigate the
connection between the complex flows found here and the quiescent
flows reported by Ji \etal \cite{ji2006} at large \Rey.

In conclusion, current laboratory experiments designed to approximate
flow profiles expected from accretion disks become turbulent at
moderate Reynolds number due to imposed boundary conditions. Although
these instabilities are generic and hence cannot possibly be avoided,
universal scaling suggests that shear mechanisms might overwhelm
endwall effects at large Reynolds number. In order to probe this
hypothesis new experiments with variable aspect ratio and different
axial boundary conditions should be conducted. These would provide
great indsight on the physical mechanisms of rotating shear flows and
might shed light on the origin of turbulence in astrophysical disks.

\begin{acknowledgments} 
  Support from the Max Planck Society is acknowledged and the
  Engineering and Physical Sciences Research Council (Grant
  No. EP/F017413/2) is acknowledged. The author is grateful to Kerstin
  Avila and Bjoern Hof for discussions.
\end{acknowledgments}

\end{document}